\documentclass[aps,preprint,superscriptaddress,showpacs,showkeys]{revtex4}

\usepackage{graphics}
\usepackage{graphicx}

\begin{document}

\title{High resolution dynamical
 mapping of social interactions with active RFID}

\author{Alain Barrat}
\affiliation{Centre de Physique Th\'eorique (CNRS UMR 6207), Marseille, France}
\affiliation{Complex Networks and Systems Group, ISI Foundation, Turin 10133, Italy}
\author{Ciro Cattuto}
\affiliation{Complex Networks and Systems Group, ISI Foundation, Turin 10133, Italy}
\author{Vittoria Colizza}
\affiliation{Complex Networks and Systems Group, ISI Foundation, Turin 10133, Italy}
\author{Jean-Fran\c{c}ois Pinton}
\affiliation{Laboratoire de Physique de l'ENS Lyon (CNRS UMR 5672), Lyon, France}
\author{Wouter Van den Broeck}
\affiliation{Complex Networks and Systems Group, ISI Foundation, Turin 10133, Italy}
\author{Alessandro Vespignani}
\affiliation{Complex Networks and Systems Group, ISI Foundation, Turin 10133, Italy}
\affiliation{School of Informatics and Biocomplexity Institute,
Indiana University, Bloomington, IN 47408, USA}

\begin{abstract}   
In this paper we present an experimental framework to gather
data on face-to-face social interactions between individuals,
with a high spatial and temporal resolution.
We use active Radio Frequency Identification (RFID) devices
that assess contacts with one another by exchanging low-power radio packets.
When individuals wear the beacons as a badge, a persistent radio contact between
the RFID devices can be used as a proxy for a social interaction between
individuals. We present the results of a pilot study 
and a subsequent preliminary data analysis,
that provides an assessment of our method and highlights
its versatility and applicability in many areas concerned with human dynamics.
\end{abstract}


\maketitle

\section{Introduction}

Social interaction patterns such as contacts and mixing patterns among individuals have
a direct impact on diverse phenomena studied in various research
areas. Clear-cut examples are the transmission of infectious diseases by 
the respiratory or close-contact route, and collective
opinion formation. The availability of representative data on such
patterns has long been a concern since it used to be notoriously
difficult to collect it. 
The available methods usually rely on surveys
and paper-diary methodologies~\cite{Mossong:2008} which are often
slow, inaccurate, and intrusive. Novel
technologies, however, afford new and promising means of collecting
this essential data.

Contact patterns data is indeed much needed. Recent studies of e-mail \cite{Ebel:2002} and
cellular phone call exchanges~\cite{Onnela:2007,Gonzales:2008},
collaboration networks~\cite{Newman:2001}, sexual contact
networks~\cite{Liljeros:2001}, and mobility by air
travel~\cite{Barrat:2004}, have revealed the presence of complex
properties and heterogeneities. In particular, the number of
interaction partners from one individual to the other is subject to 
large fluctuations that have non-trivial
consequences on the dynamical processes taking place on these
networks~\cite{Pastor:2004,Boccaletti:2006,BBVbook}. A detailed
characterization of these structures is therefore of utmost importance
for the understanding of many phenomena, and crucially depends on
the availability of representative empirical data.

While important progress has been achieved in the last decade or so,
more is bound to come, for most of these recent characterizations of
complex networks focused on static configurations in which the temporal
dimension was not considered, mostly because of lack of data. Examples
of properties that arise in this temporal dimension are duration,
frequency, concurrency, and causality. For instance, if an individual A meets first
B then C, an information or a virus can spread from B to A and
then to C, but not from C to B. In the image of a static network in
contrast, the links allow propagation in both cases. The fact that
these static networks are in fact ``summaries'' of many different
interactions that do not occur simultaneously, might conceal important
insights. The few cases in which temporal aspects have been considered
in more detail, indeed revealed important
consequences~\cite{Eckmann:2004,Barabasi:2005,Holme:2005,Vazquez:2006,Iribarren,Kostakos,Vazquez:2007,Hui:2005,Scherrer:2008}.

Several more recent studies have demonstrated the potential of using
novel technologies such as Bluetooth and Wifi for collecting data on
both the structural and temporal aspects of social interaction
patterns~\cite{Hui:2005,Eagle:2006,Kostakos,Scherrer:2008}.
However, their spatial resolution in these is at best of the order of
$10$ meters, and the temporal resolution of the order of 2-5 minutes.
Moreover, these technologies detect local proximity between devices, which does
not imply a priori a social interaction between the individuals
carrying these devices.
Finally, these studies concern small groups and are not
easily reproducible.

In this paper, we present a novel experimental
framework based on active RFID devices that overcomes these
limitations. We discuss a recently performed pilot study, and a data
analysis that highlights the main advantages of this new data collection
technique. Non-technical accounts and supplementary material can be
found on the website of the SocioPatterns project
\cite{sociopatterns}.

\section{Methodology, experiments, data}

\subsection{Active RFID-based experimental framework}

The proposed experimental framework aims at measuring the contact
patterns of a group of
interacting individuals in a spatially bounded setting, such as a
set of offices or a conference. The participants are asked to carry
small RFID tags~\cite{finkenzeller2003rh}, henceforth called
 \emph{beacons}. These beacons continuously broadcast small data
packets which are received by a number of stations and relayed
through a local network to a server. The stations are installed at
fixed locations in the environment. The beacons and stations we used
were created by and obtained from the OpenBeacon
project~\cite{openbeacon}.

RFID tags acting as beacons can be used to deploy indoors locative
systems~\cite{ni2004lil} that track the location of the tags.
Problems related to multiple path, phase fluctuations, etc. tend however to
limit the precision of the spatial localization of the tags.
Because of this, locative technologies are typically not viable, at low cost,
to infer face-to-face contact between individuals wearing RFID tags.

Moving from \emph{contact inference} to direct \emph{contact detection}
enabled us to bypass these limitations. To this end, we leveraged the
OpenBeacon active RFID platform~\cite{openbeacon} and operate the 
RFID tags a bi-directional fashion. That is, tags no longer act
as simple beacons that passively emit
signals to be received and processed by a centralized post-processing
set-up. They rather exchange messages in a peer-to-peer fashion to
sense their neighborhood and assess directly contacts with nearby tags.

A high spatial resolution of less then $1-2$ meters is attained by
using very low radio power levels for the contact
sensing. Furthermore, assuming that the subjects wear the tags on
their chest, the body effectively acts as a shield for the sensing
signals. This way, contacts are detected only when participants
actually face one another. If a sensed contact persists for a few
seconds, then given the short range and the face-to-face requirement,
it is reasonable to assume that the experiment is able to detect an
ongoing social contact (as e.g. a conversation).

After the beacons  detect a contact, they  broadcast a report
message at a higher power level. These reports are received by the
stations and relayed to the monitoring infrastructure.
The reports are stored with a time stamp, the \textsc{id} of the relaying
station and the \textsc{id} of the tags which participate in the contact
event (up to $4$ simultaneous contacts are recorded, using the current hardware).

After a suitable tuning of the system parameters, we can
easily record individual contacts in a crowded room with just a small
number of receiving stations.
The raw data series is made with an
effective sampling frequency under one second. In many instances of
data processing, we  applied a coarse graining filter using time windows of
$20$ seconds (see next section). 
This value is chosen in order to minimize
statistical errors, and corresponds moreover to a typical timescale for
social interactions.
We finally note that messages between tags
and/or stations are encrypted and that the entire data management is
completely anonymous.

\subsection{Visualization}

The pilot studies we conducted so far were accompanied with publicly
displayed, dynamic visualizations of the contacts between individuals.
This is achieved by defining a contact network in which the
beacons/persons are nodes and the contacts are edges. Two different
types of visualizations can be displayed, providing
snapshots respectively of the \emph{instantaneous} state of the network, and 
of the \emph{cumulative} state
since a given time (e.g. the start of the experiment, or the start of
the day in a multi-day experiment). The 'instantaneous' visualization
additionally displays marks for the stations, which are positioned in
a fixed layout. The location of the beacon marks in the visualization
is driven by a force-directed layout algorithm. Springs are associated
with both the explicitly shown contact edges and the edges between
beacons and stations, which are not shown. The rest length of these
springs is inversely proportional to the strength of the respective
contact or beacon-station proximity estimations.

\begin{figure}[t]
\centering
\includegraphics[width=\textwidth]{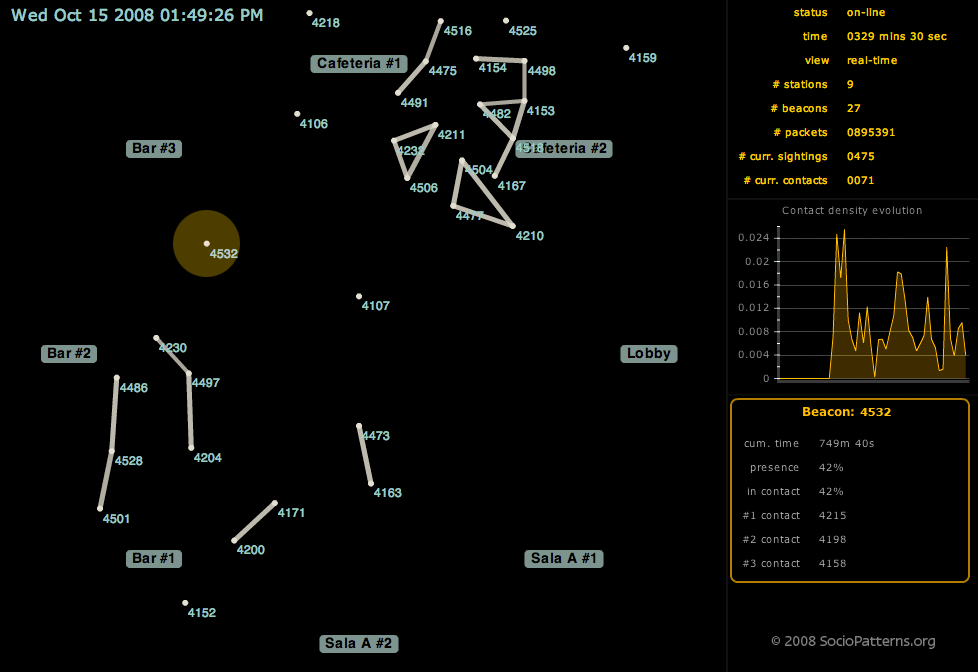}
\caption{A snapshot of the visualization. The main view shows the
instantaneous state of the contact network at a given time during
lunch hour. The beacons are labeled with their \textsc{id}s, and can
also be labeled with available metadata (such as e.g. the actual names
of the persons). White edges represent contacts. The beacons are
positioned near the stations where their signals are
received.
The yellow circle behind beacon 4532
highlights a tap while some related data is shown in the
side-bar. 
}
\label{bt2_snapshot}
\end{figure}

The model is regularly updated based on the live data feed, and the
view is updated after each iteration of the algorithm, up to 25 times
per second. The result is a continuously morphing network
representation in which the marks of beacons that are in contact try
to occupy adjacent positions, and  to move towards the
marks of the closest stations.
The other visual encodings are as follows: Edge thickness and
transparency encode contact strength; Beacon mark size encodes the
number of contacts reported by the beacon; Station mark size encodes
the number and proximity of the closest beacons.  The main network
view is furthermore flanked by a side-bar with various data points and
charts, which are dynamically updated as well.
Figure~\ref{bt2_snapshot} shows a snapshot of the
visualization. Sample movies can be viewed on the
website of the Sociopatterns project~\cite{sociopatterns}.

These visualizations were primarily developed to visually
follow and inspect the ongoing experiment and as an aid in explaining
it, but we also introduced certain affordances. One such feature
consists of enabling the participants to tap their beacon by pressing
a button. The visualization immediately reacts by highlighting the
corresponding beacon mark and temporarily showing a small table with
some detailed contextual data in the side-bar. Other affordances that
were effectively exploited by the participants are the localization of
people and the identification of an observed but unknown contact
partner of a known person.

\subsection{A pilot study}

\begin{figure}[t]
\includegraphics[width=0.4\textwidth]{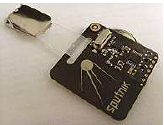}
\hspace{1cm}
\includegraphics[width=0.4\textwidth]{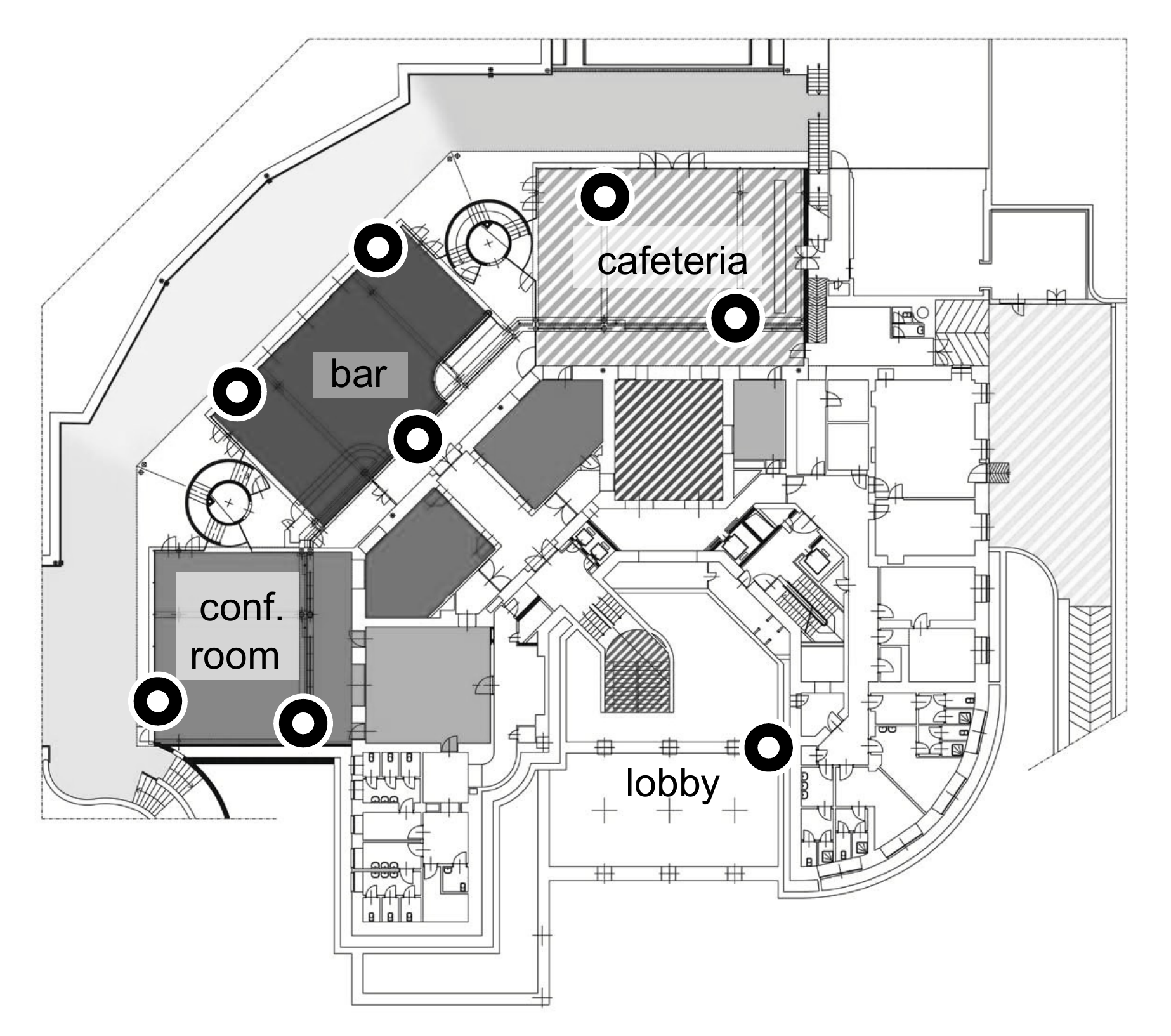}
\caption{Left: photo of a beacon (Courtesy of M. Meriac \protect\cite{openbeacon}).
Right: map of the experiment premises. The circles denote the positions of the
reporting stations.}
\label{fig:map}
\end{figure}

We have deployed our measuring infrastructure in a pilot experiment of
limited size. The experiment took place during the workshop ``Facing
the challenge of Infectious Diseases'' at the ISI Foundation on
October $13-17$, $2008$. Participants to the workshop were offered to
volunteer to participate to the experiment, and a large part agreed.
This allowed us to gather data in a very dynamical context with
periods of high social interaction (coffee and lunch breaks) and other
periods in which the participants sit together but (almost) do not
interact in a pairwise fashion. The experiment involved about $50$
attendees over four days.  We placed reporting stations in the main
areas in which people were expected to be during the sessions and
breaks~--~namely the conference room, the bar (where coffee breaks
were taking place) and the cafeteria area (where lunch was served). We
also put a station in the lobby which is also suited for discussions
(see Figure \ref{fig:map}). Figure \ref{bt2_snapshot}
presents a snapshot of the visualization obtained during the lunch break
of the third day of the conference, showing a number of beacons in the
cafeteria area where people have lunch, while others are having coffee at the bar.

The new firmware proved to be as much reliable in a real-world
setting as it appeared to be in our preliminary experiments. The measuring
infrastructure received approximately $2 \cdot 10^6$ data packets
per day from the various beacons, among which $5 \cdot 10^5$ packets
reporting a contact. Around $150$ Mb of raw compressed data were 
processed. 
Some caveats have to be reported: because of 
technical issues (some beacons had to be changed during the
experiment, some batteries failed and had to be replaced), some
beacons disappeared from the data for few hours. Moreover, 
beacons were obviously tracked only when 
within range of the stations. We will see in the next section
that, despite these issues corresponding to sampling problems,
the data analysis reveals interesting patterns and shows the large
potential of our experimental setup.

\section{Results of the pilot study}

\subsection{Contacts characterization}

\begin{figure}[t]
\includegraphics[width=\textwidth]{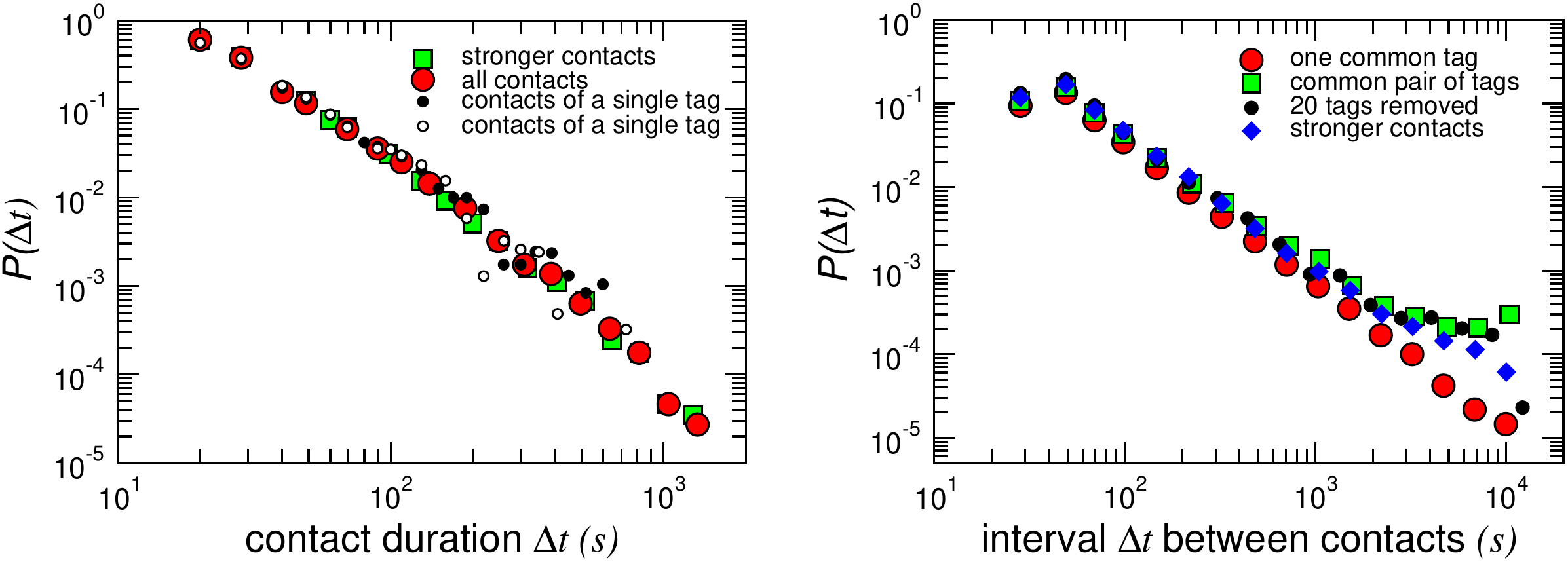}
\caption{Left: Distribution of contact durations obtained by
  considering: all contact events; the stronger contacts only
  (i.e. exchange of at least $5$ data packets between 2 beacons in a
  $20s$ time window); the contacts of two individuals selected at
  random. Right: Distribution of time intervals between contacts:
  involving at least one common beacon; involving the same pair of
  beacons; removing 20 randomly selected beacons; considering stronger
  contacts only.  }
\label{fig:durations}
\end{figure}

Let us first focus on the analysis of the contacts between individuals.
We define as a ``contact event'' between two beacons A and B the exchange of
 at least one data packet  between the two
beacons in a $20s$ time-window. We  then define as the duration of
the contact A-B the time during which packets are exchanged between them at least every
$20s$. The contact is considered as broken whenever more
than $20s$ occur without a packet exchange.
The choice of a $20s$ window is based on the frequency with which 
packets are sent by beacons, and 
corresponds to a reasonable time-scale for
social interaction (e.g. encounter, brief conversation, etc.).
Given this definition, we can measure both the duration of each
contact and the intervals between two contacts. Figure
\ref{fig:durations}(left) shows the distribution of the contact
durations obtained using the whole dataset collected during the four
conference days. A very broad distribution is observed, close to a
power-law with exponent $\simeq -2$. Qualitatively, this behavior is
not unexpected: there are comparatively few long-lasting contacts and
a multitude of brief contacts. A similar result has been reported for
the duration of contacts between Bluetooth devices
\cite{Scherrer:2008}, with different exponents depending on the
experimental set-up. Our measurements, however, achieve higher spatial
and temporal accuracy than previous studies, and reliably select
face-to-face interactions at close range, allowing to detect social
interactions of a conversational type. These measurements clearly show that
no characteristic time of interaction can be determined but that these
interactions can occur on many different timescales.

We checked the robustness of the reported behavior along several
lines. First, we verified that the distribution is the same over
different periods of time: few hours, a whole day, or the whole
conference. We also checked that it is invariant across randomly
selected groups of individuals (see Figure \ref{fig:durations}(left)),
showing that each individual has a broad distribution of contact
durations. The obtained global heterogeneity therefore stems from an
heterogeneity of the contact patterns of each individual, and not from
an heterogeneity due to the difference of behavior among individuals.
The distribution of contact durations remains unchanged, even by
assuming a stricter definition of contact. The left panel of
Figure~\ref{fig:durations} shows the result obtained by defining
stronger contacts as the ones in which at least $5$ data packets are
exchanged in a $20s$ window (instead of $1$ data packet only for the
standard definition of a contact event).

Let us now turn to the inter-contact time intervals, for which
previous studies \cite{Hui:2005,Scherrer:2008} have also reported
broad distributions. Time intervals between contacts can in fact be
defined in three different ways. One can measure the time between any
two reported contact events, regardless of the involved
beacons, thus yielding  a characterization of
the global dynamic/social activity of the group under study. We
observe a broad distribution close to a power-law with exponent $-2.5$
(not shown, see the Sociopatterns project website~\cite{sociopatterns}). 
 Different measures, which are more important in relation
with spreading processes, focus on (i) the time intervals
between two contacts  involving a given particular beacon, and (ii) 
the time intervals between two contacts involving the same pair of
beacons. Figure \ref{fig:durations}(right) displays these 
distributions, showing that also in this case a broad behavior
is obtained.  
This behavior is robust with respect to possible (heavy) data loss, as
shown by the distribution obtained by removing the data coming from $20$
randomly selected beacons that represent more than $30\%$ of the whole
dataset. The stronger contact definition
also yield similar results.

The results presented here are not unexpected, since bursty behaviors
and broad distributions of events durations or inter-event intervals
have been reported in other studies on human behavior. It is
nonetheless striking to observe that our experimental setup based on a
new technology aimed at contact detection yields high quality data
within a relatively small experiment, in agreement with the expected
behavior. We foresee that larger experiments will allow to obtain
larger statistics and to investigate more in detail the social and
dynamical aspects of contact and mixing patterns, through a detailed
characterization of the links (intermittency, persistence, etc.) and
of the nodes (role inference, inclusion of background information,
etc.).

\subsection{Social networks}

\begin{figure}[t]
\includegraphics[width=\textwidth]{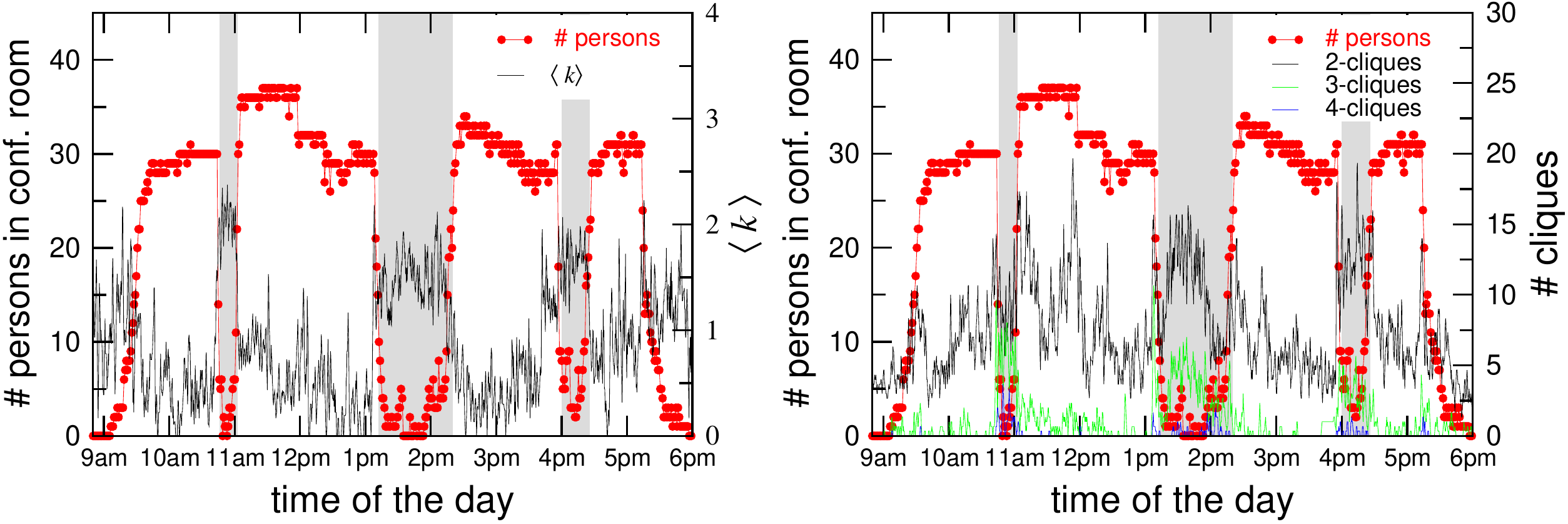}
\caption{
Number of beacons in the conference room as a function of
time, during the third day of the conference. Left: Average degree
$\langle k \rangle$ in the instantaneous contact network computed over
time windows of $20s$.  Right: Number of pairs (2-cliques), triangles
(3-cliques), and $4$-cliques in the contact network.  Note that we
consider the maximal cliques, i.e. that the three edges of a triangle
are not counted in the number of pairs.  }
\label{fig:timeline}
\end{figure}

The data on social contacts can be used to build aggregated
networks of interactions between individuals on any timescale larger
than the time resolution. Individuals are the nodes of the network, and a link 
of weight $w$ exists between two individuals if $w$ contact events have taken place between them in the
chosen time interval.
Let us first focus on ``instantaneous'' networks, constructed on short
timescales. Figure \ref{fig:timeline} shows the number of beacons in
the conference room as a function of time\footnote{Although a precise
tracking of the beacons' locations is a difficult task, it is easy to
know in which room each beacon is.},
during the third day of
the conference, which was divided into four sessions, separated by two
coffee breaks and a lunch break (indicated by the gray areas in the
Figure). The data, averaged over time windows of $20s$, clearly shows
the attendance of each session, in which most beacons are in the
conference room, whereas the breaks are identified by the small number
of beacons remaining in the conference room. The left panel also
displays the evolution of the average number of contacts per
individual during $20s$ periods. Strikingly, the number of contacts
per participant is low when the attendance in the conference room is
high, whereas a clear increase is observed during each break, clearly
signalling that most social interactions occur during the coffee and
lunch breaks, though some contacts may occur during the sessions when
people typically talk and discuss with their immediate neighbors.
This is further highlighted in the right panel of Figure
\ref{fig:timeline}, where we display, together with the attendance
curve in the conference room, the number of $2-$, $3-$ and $4-$
cliques in the contact networks aggregated over $20s$ time
windows. Note that we consider here maximal cliques, so that the edges
of a triangle are not counted as $2-$cliques, or that the $4$
triangles forming a $4-$clique are not counted in the number of
$3$-cliques. A fluctuating number of pairs is observed during the
session, corresponding most probably to participants turning towards
their neighbours, and peaks are observed at the beginning and end of
each session and in fact of each talk, when participants have indeed
more activity. $3-$ and $4-$ cliques are observed almost exclusively
during the breaks, as expected since many discussions take place in
small groups. It is worth to mention that the small number of
$3-$cliques observed during the sessions correspond to small groups of
participants remaining in the coffee break area for discussions even
after the beginning of the session.

The results illustrated in Figure~\ref{fig:timeline} are clearly
expected, since social interactions obviously take place during the
breaks. However, they point to the ability of our experimental setup
of resolving the mixing patterns by directly detecting the contact
events. A less elaborate setup, based on the inference of contact
events by spatial proximity, would show a large number of cliques (or
worse, a unique large clique) during the meeting session where
participants are physically close. In addition,
Figure~\ref{fig:timeline}(right) clearly shows how this technology is
able to detect interactions between $3$ or $4$ people, and not only
pairwise interactions.

\begin{figure}[t]
\includegraphics[width=\textwidth]{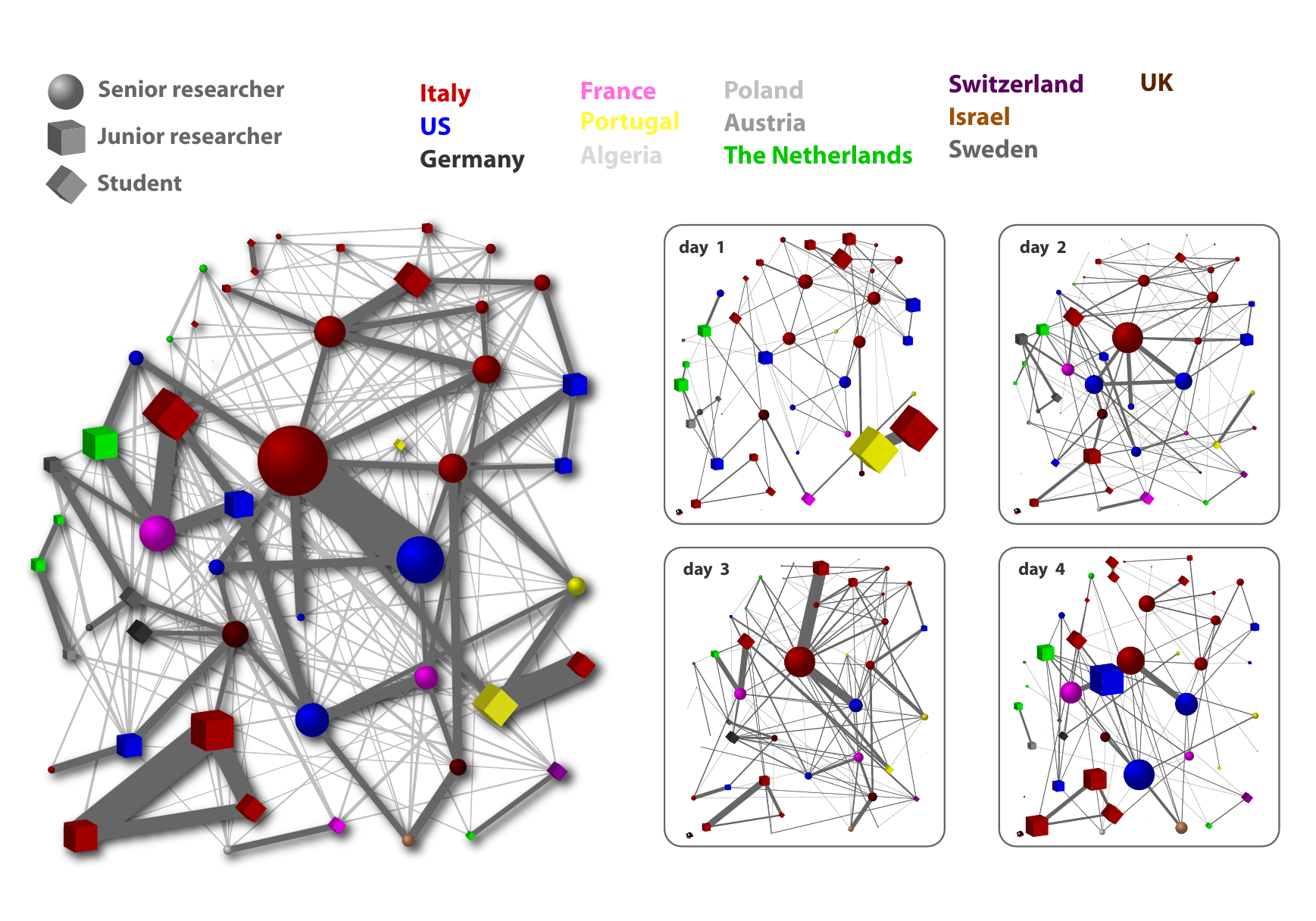}
\caption{Social network of contacts between individuals (represented
  by nodes), aggregated over the whole duration of the conference
  (larger graph), and for each of the days (smaller panels).  The size
  of each node is proportional to its strength (given by the sum of
  the weights of its links \protect\cite{Barrat:2004}), and the width
  of each link is proportional to its weight. The color of each node
  corresponds to the individual's country of affiliation, and the
  shape to his/her academic position.  For clarity, only links with
  weight larger than 100 are reported (50 for the smaller panels). As
  visible from the smaller panels, different interaction patterns are
  obtained for different days.  }
\label{fig:net}
\end{figure}

The data can also be used to construct aggregated networks on longer
timescales, for example for a single day or for the whole duration of
the experiment. The aggregated network becomes then denser as the
aggregation time increases, with an average degree ranging from a
value close to $20$ for the network aggregated over one day, to
approximately $40$ for the whole experiment duration, showing that
most participants have interacted with each other, which is in fact
one of the aims of a small-scale conference. The aggregated networks
are interesting in that they show broad distributions of the weights
(given by the number of packets exchanged between two beacons) which
are a proxy for the effective duration of a social
interaction. Without going into a detailed network analysis, we
provide in Figure~\ref{fig:net} a visualization of the networks of
social interactions obtained by aggregating the data for each day of
the conference (smaller panels) and for its entire duration (larger
graph), with the heterogeneity of links weights and nodes strengths
clearly visible.

\subsection{Contagion processes}

\begin{figure}[b!]
\includegraphics[width=\textwidth]{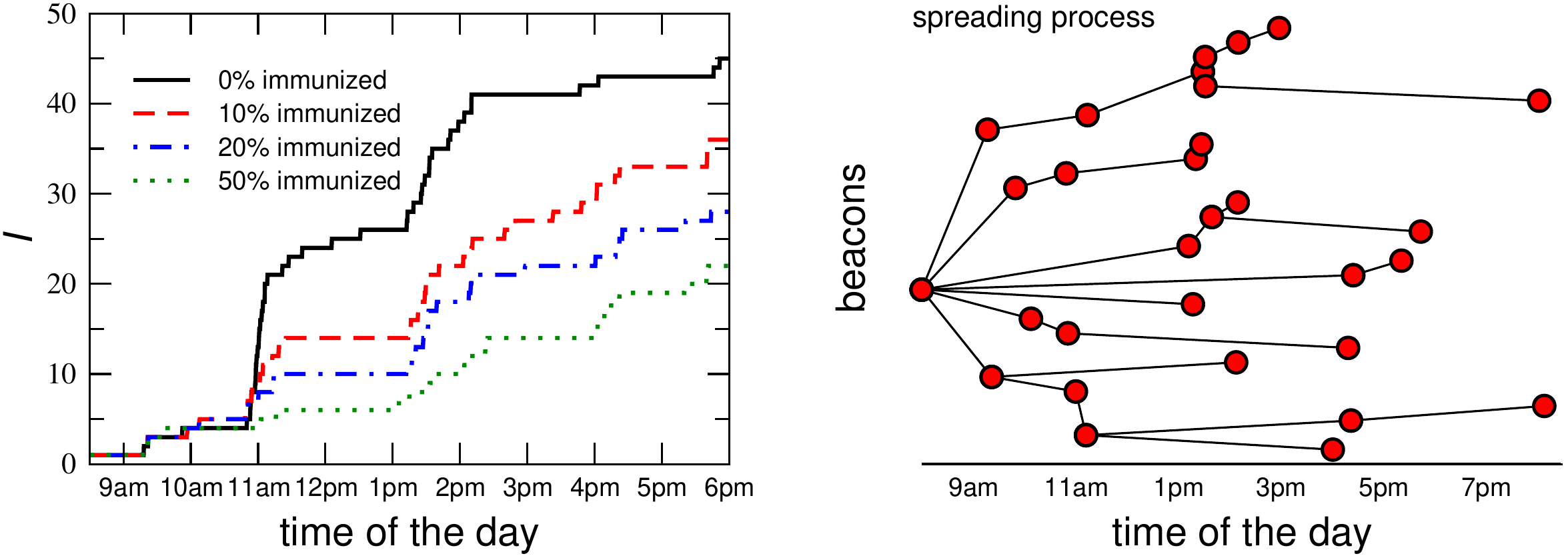}
\caption{Left: Evolution of the number of 'Infected' individuals when a
single Infectious is introduced at the beginning of the day. For each
contact, the transmission probability is $0.01w$ for
each $20s$ time window, where $w$ is the number of packets exchanged
between the two beacons in contact during this time window. Right:
illustration of the contagion events in the population of beacons as a function 
of time, for $20\%$ initially immune individuals. Black lines 
indicate the infection from one beacon to another, as they occur in time.
}
\label{fig:spread}
\end{figure}

The dynamic network of contacts provides a realistic setting to
perform simulations of contagion processes in the population of
individuals, such as rumour or information spreading,
opinion formation, or epidemic processes. Particularly relevant is
the application to the spread of infectious diseases transmitted by
the respiratory or close-contact route (as for example influenza,
SARS, etc.). Models of epidemic spread on contact networks usually
rely on static configurations of networks where the aspects of
concurrency and causality are not taken into account. The data
collected with our experimental setup can be used for an emulation of
a contagion process among individuals where all topological and
temporal heterogeneities are considered.

Here we present a very simple example of a contagion process aimed at
 showing the feasibility of such studies. We consider the basic
 Susceptible-Infected (SI) model in which individuals are classified
 in two mutually exclusive compartments, Susceptible (i.e. able to
 contract a disease) and Infectious (i.e. infected and able to
 transmit the infection)~\cite{Maybook}.  The emulation is performed
 on the contact data of the third conference day. At the beginning of
 the day, a randomly selected individual is considered as infectious.
 During each time window of $20s$, each contact between a susceptible
 and an infectious individual can result in the contagion of the
 susceptible that contracts the infection with probability
 $0.01w$ (where $w$ is the number of packets exchanged between
 the beacons of the individuals, i.e. a measure of the intensity and
 duration of the social interaction). Some individuals are set as
 immune since the start of the emulation, allowing for individuals who
 are not susceptible to the disease and can never become infectious.
 Figure~\ref{fig:spread} displays the number of infectious individuals
 as a function of time for a single realization of the stochastic
 model, and for different percentages of initially immune
 individuals. An interesting pattern is observed, in agreement with
 the previous analysis: most contagion events occur during the coffee
 and lunch breaks, where social interactions are more likely to
 occur. The right panel displays a schematic visualization of the
 propagation dynamics, shown as a tree in which each newly
infected beacon is represented as a red disk at the time of its contamination, with lines going from
the infecting beacon to the infected one for each contagion event.
While this model is overly simplistic and does not aim to reproduce a
given realistic epidemic scenario, it offers the possibility of
studying simple contagion processes on a realistic dataset, and
provides a proof of concept showing how the data gathered through our
experimental set-up in proper settings (as e.g. larger social events)
can have a crucial value to understand and predict the impact of
infectious diseases.

\section{Conclusions and perspectives}

In this paper, we  presented a novel experimental set-up which can
be used to gather information on social interactions of
individuals. The measures are based on active RFID devices, called
beacons, that individuals can wear 
as badges.  When two beacons are close enough
(typically one meter apart), they can exchange messages and relay them
to the measuring infrastructure. The very low power used for the
exchanged messages and the absorption of the used frequencies by 
the human body  ensure that contacts are detected only when
individuals face each other as in a real social contact. This allows us
to obtain data at very high spatial and temporal resolution, as
shown in a pilot experiment performed during a recent
conference. Here we  presented some results of the corresponding
data analysis, showing the resolving power of experimental setup,  able to
discriminate between social interaction and simple physical proximity.
We  measured the distributions of the duration of social contacts
between individuals and of the intervals between contacts,
and found broad behaviors. Moreover, we showed how our experimental
setup can be used to construct social networks by aggregating the
contacts over the required timescale. 

Our experimental set-up paves the way for a number of developments and
applications. Clearly, more experimental work is
needed
to obtain 
larger statistics on contacts durations or frequencies, and to characterize
dynamically evolving social networks. The hardware and
software could also be upgraded to contain additional information 
on the individuals and their interactions. 

The presented set-up will also allow to study various dynamical
phenomena taking place on dynamically evolving contact networks,
as briefly illustrated above. Contagion processes,
such as rumour spreading, opinion formation, propagation of respiratory 
or close-contact infections, take place on the dynamical network of 
social contacts among individuals. Gathering data on social contacts 
will allow a better modeling and understanding of the spread of viruses and information.

\vspace*{0.1cm}
\noindent
\small{V.C. is partially funded by the European Commission contract
  n. ERC--2007--Stg204863 (EpiFor).  A.V. is partially funded by the
  NIH-NIDA-21DA024259-01 award.}

\end{document}